\documentclass[aps,prd,showpacs,floatfix,twocolumn]{revtex4}

\def\be{\begin{equation}}
\def\ee{\end{equation}}

\DeclareMathSymbol{\varGamma}{\mathord}{letters}{"00}

\begin{document}

\title{$X(3872)$ as a $^1D_2$ charmonium state}

\author{Yu. S. Kalashnikova}

\affiliation{Institute for Theoretical and Experimental Physics, 117218,
B.Cheremushkinskaya 25, Moscow, Russia}

\author{A. V. Nefediev}

\affiliation{Institute for Theoretical and Experimental Physics, 117218,
B.Cheremushkinskaya 25, Moscow, Russia}

\begin{abstract}
The $^1D_2$ charmonium assignment for the $X(3872)$ meson is considered, as prompted by a recent result from the BABAR Collaboration, favouring $2^{-+}$ quantum numbers for the $X$. It is shown that established properties of the $X(3872)$ are in a drastic conflict with the $^1D_2$ $c\bar{c}$ assignment.
\end{abstract}


\maketitle

Seven years after its discovery, the $X(3872)$ meson has confirmed once again its
reputation of {\it enfante terrible} of meson spectroscopy. The state was first seen by Belle \cite{Xobservation} in the $\pi^+\pi^-J/\psi$ mode and then confirmed, in the same discovery mode, by the CDF \cite{CDF}, D$\emptyset$ \cite{D0}, and BABAR \cite{BABAR} Collaborations. According to the CDF analysis of the dipion mass spectrum and the angular distribution in the $ \pi^+\pi^-J/\psi$ mode \cite{CDFjpc}, only $1^{++}$ and $2^{-+}$ assignments are able to describe the data. Then, while the nature of the state remains controversial, there are good phenomenological reasons to assign it $1^{++}$ quantum numbers. 

First, the $X$ resides at the $D^0 \bar{D}^{*0}$ threshold, which prompts a considerable admixture of a molecule in its wave function. 
Furthermore, CDF concludes that the $\pi^+\pi^-$ come from the $\rho$ \cite{rho} which, together with the
Belle observation of the $\omega J/\psi$ mode \cite{Xomega}, points to a considerable isospin violation. The latter can be explained naturally in the molecular model of the $X$, which implies $1^{++}$ quantum numbers. 
In addition, the $X$ was also observed in the $D^0 \bar{D}^{*0}$ mode with a significant rate \cite{Belleddpi,Babarddpi,Belle3875v2}. Both $\rho J/\psi$ and $D^0 \bar{D}^{*0}$ modes were analysed simultaneously in Refs.~\cite{recon,xnew,braaten}, and it was shown that indeed the data were compatible with a large admixture of the $D^0 \bar{D}^{*0}$ molecular component in the wave function of the $X$. 

However, a recent analysis of the decay $B \to K \omega J/\psi$ data performed by the BABAR Collaboration \cite{babarjpc} indicates that inclusion of an extra unit of the orbital angular momentum in the $\omega J/\psi$ system improves significantly the overall description of the observed $\pi^+\pi^-\pi^0$ mass distribution, which implies a negative $P$-parity of the $X(3872)$ state. Although this new BABAR result is fully compatible with the $2^{-+}$ assignment allowed by CDF, if confirmed, it clearly challenges our understanding of the charmonium spectroscopy above the open-charm threshold. Here we investigate the most conventional explanation for the $2^{-+}$ $X(3872)$ as the $1^1D_2$ charmonium state.

In case of charmonium $D$-levels we have an experimental anchor at our disposal --- the $\psi(3770)$ vector state which is dominantly a $c\bar{c}$ state, with the angular momentum of the quark--antiquark pair $L=2$ and the total quark spin $S=1$ ($^1D_2$ state has $L=2$ and $S=0$). As $c$-quark is heavy, the spin--orbit force, which splits spin-triplet and spin-singlet levels, is not large and all $D$-levels are degenerate in the leading-order approximation. Hence, one may use the data on the $^3D_1$ level to estimate the mass and matrix elements of the $^1D_2$ level.

It became clear long ago that the $^1D_2$ assignment for the $X$ disagreed with quark model mass estimates (see, for example, Refs.~\cite{BG,alla,hch}). Indeed, quark models usually predict the $1^1D_2$ mass in the range $3770\div 3830$~MeV, while the mass difference between the $1^1D_2$ and $1^3D_1$ levels is predicted to be, averagely, $20\div 30$~MeV. Thus quark models cannot accommodate the $X(3872)$ as a
$^1D_2$ state. The same conclusion was drawn in a recent paper \cite{burns}. One might think that inclusion of various $D$-meson loops changes this statement. It is not the case, however. Loop calculations 
in the Cornell decay model \cite{eichten} and in the $^3P_0$ decay model \cite{yusk} give for the mass of the $1^1D_2$ level $3838$ MeV and $3800$~MeV, respectively. 

The arguments based solely on the mass calculations are, of course, not enough to rule out the charmonium assignment for the $X$. However further reasons for the $^1D_2$ interpretation to be problematic have started to show up. First, radiative decay transitions $^1D_2\to\gamma J/\psi(\psi')$ rates are shown to be incompatible with the data \cite{chinese}. Second, the production cross section of the $^1D_2$ level at CDF is predicted to be much smaller than the one actually observed for the $X$ \cite{burns}. In this paper we identify a couple of new problems. Namely, we expand on the issue of radiative decays and discuss the $D^0 \bar{D}^{*0}$ mode of the $X$.

The BABAR Collaboration has reported the following rates for the decays
$X(3872)\to\gamma J/\psi(\psi'(3686))$ \cite{babargamma}:
\begin{eqnarray}
{\cal B}_1&=&Br(B^{\pm} \to K^{\pm} X)Br(X \to \gamma J/\psi)\nonumber\\
&=&(2.8 \pm 0.8 \pm 0.2) \times 10^{-6},\nonumber\\[-2mm]
\label{babargamma}\\[-2mm]
{\cal B}_2&=&Br(B^{\pm} \to K^{\pm} X)Br(X \to \gamma \psi')\nonumber\\
&=&(9.5 \pm 2.9 \pm 0.6) \times 10^{-6}.\nonumber
\end{eqnarray}
In the meantime, the upper limit on the total branching fraction $B \to K X(3872)$ imposed by BABAR \cite{total} is
\be
{\cal B}_{tot}=Br(B \to K X) < 3.2 \times 10^{-4}.
\label{babartotal}
\ee

Below we demonstrate that measurements (\ref{babargamma}) and (\ref{babartotal}) cannot be reconciled with each other under the assumption of the $X$ being a $1^1D_2$ charmonium. To this end we notice that the leading multipole for the $^1D_2 \to \gamma V$ ($V$ is a vector charmonium) transition is $M1$, with the width given by a standard formula (see, for example, Ref.~\cite{hch}):
\begin{eqnarray}
\varGamma\left(^{2S+1}L_J \right.&\to&\left. {}^{2S'+1}L'_{J'}\right)\label{M1}\\
&=&\frac{4}{3}~\frac{2J'+1}{2L+1}\delta_{LL'}
\delta_{S,S' \pm 1}\frac{\alpha e_c^2}{m_c^2}|\left<\psi_f|\psi_i\right>|^2E_{\gamma}^3,\nonumber
\end{eqnarray}
where $m_c$ is the charmed quark mass, $e_c=2/3$, $\psi_i(\psi_f)$ is the initial(final)-state radial wave function, and $E_{\gamma}$ is the photon energy. In this formalism, the transition $^1D_2\to\gamma J/\psi(\psi')$ is a so-called hindered transition, so that $\left<\psi_f|\psi_i\right>=\sin\theta$, where $\theta$ is the $^3S_1-{}^3D_1$ mixing angle.
Thus the amplitude simply vanishes if $J/\psi(\psi')$ is assumed to be a pure $^3S_1$ state. The standard value for the $\psi'$ is $\theta\approx 12^0$, which gives (for $m_c=1.5$~GeV):  
\be
\varGamma\left(^1D_2(3872) \to \gamma \psi'\right)\approx 6.6[\mbox{keV}]\sin^2\theta\approx 0.29~\mbox{keV}.
\label{3686}
\ee
Notice that, being almost a pure $^3S_1$ state, $J/\psi$ possesses a tiny mixing angle $\theta$, so that even a much larger photon energy ($E_{\gamma}=698$~MeV for the $\gamma J/\psi$ final state versus $E_{\gamma}=186$~MeV for $\gamma\psi'$) cannot provide a sizable contribution of this, formally leading, $M1$ transition. Therefore, contributions of higher multipoles have to be considered. In Ref.~\cite{chinese} both widths were calculated in a quite elaborated (though rather model-dependent) NRQCD approach, with the result:
\begin{eqnarray}
&&\varGamma\left(^1D_2(3872)\to\gamma \psi'\right)\approx 0.45\div 0.5~\mbox{keV},\label{3686ch}\\
&&\varGamma\left(^1D_2(3872)\to\gamma J/\psi\right)\approx 6.8\div 9.5~\mbox{keV},\label{jpsich}
\end{eqnarray}
claimed in Ref.~\cite{chinese} to contradict the BABAR data (\ref{babargamma}).

For the case of the $\psi'$, the $M1$ contribution from the $^3S_1-{}^3D_1$ mixing to the result (\ref{3686ch}) is indeed dominant and it is in a good agreement with the simple estimate (\ref{3686}). Notice that formula (\ref{M1}) does not take into account recoil corrections, while the formalism of Ref.~\cite{chinese} accounts for the recoil only via the multipole expansion. Because of a small photon energy this seems reasonable for the $\psi'(3686)$ final state while, for the $J/\psi$ final state, the photon energy is much larger, so that the value (\ref{jpsich}) is probably an overestimation.

In order to estimate the total width of the $^1D_2$ charmonium we notice that, as is well-known 
(see, for example, Refs.~\cite{BG,hch,fazio}), the main radiative transition of the $^1D_2$ state is
$^1D_2\to\gamma ^1P_1(3525)$ with the width:
\be
\varGamma\left(^1D_2(3872)\to\gamma ^1P_1(3525)\right)\approx 460(345)~\mbox{keV~\cite{BG}(\cite{fazio})}.
\label{e1width}
\ee

Alternatively, this width can be estimated from the measured branching fractions for the transitions \cite{PDG}:
\begin{eqnarray*}
&&Br\left(\psi(3770) \to \gamma ^3P_0(3415)\right)=(7.3 \pm 0.9) \times 10^{-3},\\
&&Br\left(\psi(3770) \to \gamma ^3P_1(3510)\right)=(2.9 \pm 0.6)\times 10^{-3}.
\end{eqnarray*}
Indeed, the leading multipole is $E1$, with the width:
$$
\varGamma\left(^{2S+1}L_J \to{}^{2S'+1}L'_{J'}\right)=\frac{4}{3}C_{fi}\delta_{SS'}e^2_c\alpha|
\left<\psi_f|r|\psi_i\right>|^2E^3_{\gamma},
$$
where $C_{fi}=\mbox{max}(L,L')(2J'+1)\left\{
\begin{array}{ccc}
L'&J'&S\\
J&L&1\\
\end{array}
\right\}^2$. In the heavy-quark limit, the dipole matrix element $\left<\psi_f|r|\psi_i\right>$ is the same for all $D \to P$ transitions. This gives
\be
\varGamma\left(^1D_2(3872\right) \to \gamma ^1P_1(3525))\approx 340\div 440~\mbox{keV},
\ee
in a good agreement with Eq.~(\ref{e1width}). 

Hadronic modes of the $^1D_2$ charmonium were estimated in Ref.~\cite{BG}. These are the light hadrons (``$gg$") modes
and the $\eta_c\pi\pi$ mode:
\begin{eqnarray*}
&&\varGamma\left(^1D_2(3872)\to\mbox{light~hadrons}\right)\approx 190~\mbox{keV},\\
&&\varGamma\left(^1D_2(3872)\to\eta_c \pi\pi\right)\approx 210 \pm 110~\mbox{keV}.
\end{eqnarray*}
Together with the radiative decay modes these give for the total width the value $\varGamma_{tot}\approx 800~\mbox{keV}$.
In principle, $\varGamma_{tot}$ should also include a contribution of the $D\bar{D}^*$ modes. However, as will be shown below, with the suppression of the $D^0 \bar{D}^{*0}$ mode, this contribution is negligible.

Therefore, if we use, in accordance with the Ref.~\cite{chinese} calculation, the value of about 8~keV for the 
$\gamma J/\psi$ width, we get:
\be
Br(B \to K X)=(800/8)\times{\cal B}_1=(2\div 3)\times 10^{-4},
\label{psitotal}
\ee
which is compatible with Eq.~(\ref{babartotal}). In the meantime, a similar estimate for the decay $X\to\gamma\psi'$ gives:
\be
Br(B\to KX)=(800/0.5)\times{\cal B}_2=16\times 10^{-3},
\label{2stotal}
\ee
where $\varGamma(X\to\gamma\psi')=0.5$ keV as per (\ref{3686ch}) was used. This value is awfully larger than the upper limit (\ref{babartotal}). To reconcile ${\cal B}_2$ with ${\cal B}_{tot}$ one needs to decrease $\varGamma_{tot}$ fifty times.

We conclude in such a way that the data on the radiative decays of the $X(3872)$ do not allow for its $^1D_2$ charmonium interpretation, if the BABAR result on the $\gamma\psi'$ mode holds true. Notice, however, that the most recent Belle results \cite{bellegamma} read:
\begin{eqnarray*}
Br(B^{\pm} \to K^{\pm} X)Br(X &\to& \gamma J/\psi)\\
&=&(1.78^{+0.48}_{-0.44}\pm 0.12) \times 10^{-6},\\
Br(B^{\pm} \to K^{\pm} X)Br(X &\to& \gamma \psi')<3.4 \times 10^{-6},
\end{eqnarray*}
which suggests that the BABAR and Belle measurements for the $\gamma\psi'$ mode contradict each other.

Let us now consider the $D^0 \bar{D}^0 \pi^0$ mode.
In 2006 the Belle Collaboration reported an enhancement of the 
$D^0\bar{D}^0\pi^0$ signal observed in the reaction $B^+\to K^+D^0\bar{D}^0\pi^0$ just above the $D^0\bar{D}^{*0}$  threshold \cite{Belleddpi}, at $M_X=3875.2\pm 0.7^{+0.3}_{-1.6}\pm 0.8~\mbox{MeV}$, with the branching
$$
Br(B^+\to K^+D^0\bar{D}^0\pi^0)=(1.02\pm 0.31^{+0.21}_{-0.29})\times 10^{-4}.
$$
The peak was confirmed by the BABAR Collaboration as well \cite{Babarddpi}.
However, recently the Belle Collaboration announced a new analysis for the 
$D^{*0}\bar{D}^0$ case \cite{Belle3875v2}, and a lower peak position was obtained than reported before, namely, $M_X= 3872.9^{+0.6+0.4}_{-0.4-0.5}~\mbox{MeV}$, with the branching
\be 
Br(B^+\to K^+D^0\bar{D}^{*0})=(0.8\pm 0.2 \pm 0.1)\times 10^{-4}.
\label{BrBnew}
\ee
This enhancement was associated with the $X(3872)$ state seen in the $D^0\bar{D}^{*0}$ mode (here and in what follows an obvious shorthand notation $D^0\bar{D}^{*0}\equiv D^0\bar{D}^{*0}+\bar D^0 D^{*0}$ is used), and the mass shift was attributed to the proximity to the $D^0\bar{D}^{*0}$ threshold. 

To proceed we find the ratio of branching fractions:
\be
R=Br(B \to KD^0 \bar{D}^{*0})/Br(B \to K X)>0.25,
\label{r}
\ee
where the lower limit for $R$ was deduced from the data quoted in Eqs.~(\ref{babartotal}) and (\ref{BrBnew}).
In what follows we argue that it is not possible to reproduce such a large value of the ratio $R$ under the assumption of the $X$ being the $1^1D_2$ charmonium. Indeed, it is claimed in Ref.~\cite{dunwoodie} that the peak position in the $D\bar{D}^*$ invariant mass depends on the orbital momentum $l$ of the $D\bar{D}^*$ pair. In particular, it is shown that with $l=1$ it is quite easy to produce a peak at about $3$~MeV above the $D\bar{D}^*$ threshold, accommodating in such a way both BABAR \cite{Babarddpi} and old Belle \cite{Belleddpi} measurements. Depending on the model parameters, a peak much closer to the threshold can also be reproduced with $l=1$, so that there is no contradiction with the new Belle data \cite{Belle3875v2} either.
The value $l=1$ corresponds to the $2^{-+}$ quantum numbers of the $X$ and therefore suggests the $^1D_2$ assignment for the latter. However, then the $D^0\bar{D}^{*0}$ rate behaves as $k^3$ ($k$ being the relative momentum in the $D^0\bar{D}^{*0}$ system), so the proximity to the $D^0\bar{D}^{*0}$ threshold implies a considerable suppression of the production rate. Below we make this argument quantitative.

The ratio $R$ can be calculated as
\begin{widetext}
\be
R=\frac{\displaystyle \int^{M_+}_{M_-} dM \left(dBr(B\to K D^0 \bar{D}^{*0})/dM\right)}
{\displaystyle\int^{M_+}_{M_-} dM \left(dBr(B\to K{\rm non}(D^0 \bar{D}^{*0})/dM\right)+
\int^{M_+}_{M_-} dM \left(dBr(B\to KD^0\bar{D}^{*0})/dM\right)},
\label{rexp}
\ee
where the integration takes place over the mass region where the $X(3872)$ resides, conveniently defined as
$M_{\pm}=M_0 \pm 10$ MeV, with $M_0$ being the $X(3872)$ mass.
The $D^0\bar{D}^{*0}$ and non$(D^0 \bar{D}^{*0})$ rates entering expression (\ref{rexp}) are:
\be
\frac{dBr(B \to K D^0 \bar{D}^{*0})}{dM}=\frac{{\cal B}}{2\pi}\frac{g\left(^1D_2 \to D^0 \bar{D}^{*0}\right)k^3}{(M-M_0)^2+\varGamma_{tot}^2/4},
\quad
\frac{dBr(B\to K {\rm non}(D^0 \bar{D}^{*0}))}{dM}=\frac{{\cal B}}{2\pi}\frac{\varGamma({\rm non}(D^0 \bar{D}^{*0}))}
{(M-M_0)^2+\varGamma_{tot}^2/4},
\ee
\end{widetext}
where $M_{th}=m(D^0\bar{D}^{*0})$, and the constant ${\cal B}$ absorbs the details of the short-ranged dynamics of the $b$-quark decay. Due to the factor $k^3$ the expression for the $D^0\bar{D}^{*0}$ does not take a Breit--Wigner form. 
To account for the finite width of the $D^{*0}$ we assume for $k(M)$ a simple ansatz \cite{fw} 
$k(M)=\sqrt{\mu}\sqrt{\sqrt{(M-M_{th})^2+\varGamma_*^2/4}+(M-M_{th})}$,
where $\mu$ is the $D^0 \bar{D}^{*0}$ reduced mass, $\varGamma_* \approx 65$~keV is the width of the $D^{*0}$ meson estimated from the data \cite{PDG} on the $D^{*\pm}$ meson. The standard expression for the two-body relative momentum is readily reproduced as $\varGamma_* \to 0$.
Finally, anticipating a strong suppression of the $D^0 \bar{D}^{*0}$ mode, we substitute 
$\varGamma({\rm non}(D^0 \bar{D}^{*0}))\approx\varGamma_{tot}$.

The coupling $g\left(^1D_2 \to D^0 \bar{D}^{*0}\right)$ can be estimated in the $^1D_2$ model for the $X$ using the $^3D_1$ state $\psi(3770)$ as a benchmark ($p_{DD}$ is the relative $D\bar{D}$ momentum and the charged--neutral meson mass difference is neglected):
\be
\varGamma\left(^3D_1 \to D \bar{D}\right)=g\left(^3D_1 \to D \bar{D}\right)p_{DD}^3.
\ee

We now invoke the ``loop theorems" proven in Ref.~\cite{loops}. In particular, it is shown in this paper that, in the heavy-quark limit, strong open-flavour total widths for the states in a given $\{NL\}$ multiplet ($N$ is the radial quantum number while $L$ is the quark--antiquark orbital angular momentum) are equal. The heavy-quark limit implies
that (i) the initial states are degenerate in mass and have the same wave functions within a given multiplet and (ii) the final two-meson states exhibit the same degeneracy. The decay model should satisfy some general conditions listed in Ref.~\cite{loops} (for example, the popular $^3P_0$ pair creation model satisfies these conditions, and so does the Cornell decay model).

Specifically, in the ideal heavy-quark world, the masses of all $1D$ states are identical, and the masses of the final-state $D$ and $D^*$ mesons are identical too. The partial widths into certain $D^{(*)}\bar{D}^{(*)}$ channels depend on quantum numbers of a given initial state, while the sum of partial widths over all possible 
$D^{(*)} \bar{D}^{(*)}$ final states is the same within a given $1D$ multiplet. 
In the real world, if the quark--antiquark pair in the initial meson is heavy, the theorem is violated mainly by spin-dependent interactions, which remove the mass degeneracy both in the initial and final states. One may write therefore:
\begin{eqnarray*}
&&g\left(^1D_2 \to D^0 \bar{D}^{*0} \right)=g_0\left|C\left(^1D_2\right)\right|^2,\\
&&g\left(^3D_1 \to D \bar{D}\right)=g_0\left|C\left(^3D_1\right)\right|^2,
\end{eqnarray*}
where $g_0$ is the coupling constant
common for all members of the $1D$ multiplet, while $C\left(^1D_2\right)$ and $C\left(^3D_1\right)$ are the spin--orbit recoupling coefficients for the $^1D_2\to D^0\bar{D}^{*0}$ and $^3D_1\to D\bar{D}$ decays, respectively (notice that both charged and neutral $D \bar{D}$ channels contribute to the coefficient $C\left(^3D_1\right)$, while only the $D^0 \bar{D}^{*0}$ channel contributes to the coefficient $C\left(^1D_2\right)$).
These spin--orbit recoupling coefficients were calculated in the Cornell decay model (see Table II of Ref.~\cite{eichten}) and in the $^3P_0$ decay model (see Table IV of Ref.~\cite{yusk}). Both models yield $\left|C\left(^1D_2\right)\right|^2=\frac{3}{5}\left|C\left(^3D_1\right)\right|^2$,
so that, since the $D\bar{D}$ mode is dominant for the $\psi(3770)$, we estimate the coupling $g\left(^1D_2 \to D^0 \bar{D}^{*0}\right)$ as:
\be
g\left(^1D_2 \to D^0 \bar{D}^{*0}\right)\approx\frac{3}{5}\frac{\varGamma(\psi(3770))}{p_{DD}^3}.
\ee

\begin{table}[t]
\caption{The ratio $R$ (see Eq.~(\ref{r})) for various values of the $X(3872)$ mass $M_0$ and the total width $\varGamma_{tot}$.}
\label{t1}
\begin{ruledtabular}
\begin{tabular}{c|ccc}
$M_0$, MeV&$\varGamma_{tot}=200$ keV&$\varGamma_{tot}=800$ keV&$\varGamma_{tot}=3200$ keV\\
\hline
3870.8&0.023&0.023&0.022\\
3871.4&0.033&0.032&0.029\\
3872.0&0.073&0.052&0.038\\
\end{tabular}
\end{ruledtabular}
\end{table}

In Table~\ref{t1}, we list the results for the ratio $R$ for several values of $\varGamma_{tot}$. The masses used are $m(D^0)=1864.84$~MeV and 
$m(D^{*0})=2006.96$~MeV. The $\psi(3770)$ width is $23$~MeV. Finally, for the mass $M_0$ we take the same values as used in Ref.~\cite{dunwoodie}. As described above, $\varGamma_{tot}=800$~keV is our preferred value. We have also calculated the ratio $R$ for $\varGamma_{tot}$ four times smaller as well as four times larger than 800~keV, the latter value being a bit larger than 2.3~MeV quoted in PDG \cite{PDG} as the upper limit for the width of the $X$.
Clearly all values of $R$ listed in the Table~\ref{t1} are far too low in comparison with the value (\ref{r}) deduced from the data. In addition we confirm the $D^0 \bar{D}^{*0}$ lineshapes obtained in Ref.~\cite{dunwoodie},
however, the rate appears to be quite small.
Thus we conclude that the data on the $D^0 \bar{D}^0 \pi^0$ mode contradict the $^1D_2$ charmonium interpretation of the $X$. 

To summarize, we have shown that the $1^1D_2$ charmonium assignment for the $X(3872)$ meson contradicts the existing data on its radiative decays and its $D^0 \bar{D}^0 \pi^0$ mode. Our study does not challenge the $2^{-+}$ quantum numbers. We
rather claim that, if the aforementioned experimental data are taken as a true guide, the conventional charmonium model is not able to accommodate for the $2^{-+}$ $X(3872)$. If the BABAR result on the quantum numbers of the $X(3872)$ persists, it would mean that some kind of a new interloper enters the game.

\begin{acknowledgments}
We acknowledge useful discussions with P.~Pakhlov and
support by the State Corporation of Russian
Federation ``Rosatom'', RFBR (grants \# 09-02-91342-NNIOa and 09-02-00629a), DFG (grant \# 436 RUS 113/991/0-1(R)), FCT 
(grant \# PTDC/FIS/70843/2006-Fi\-si\-ca), and ``Dynasty'' foundation and ICFPM.
\end{acknowledgments}


\begin{thebibliography}{99}
\bibitem{Xobservation} S.-K. Choi {\it et al.} [Belle Collaboration], Phys. Rev. Lett. {\bf 91}, 262001 (2003).
\bibitem{CDF} D. Acosta {\it et al.} [CDF II Collaboration], Phys. Rev. Lett. {\bf 93}, 072001 (2004).
\bibitem{D0} V. M. Abazov {\it et al.} [D$\emptyset$ Collaboration], Phys. Rev. Lett. {\bf 93}, 162002 (2004).
\bibitem{BABAR} B. Aubert {\it et al.} [BABAR Collaboration], Phys. Rev. D {\bf 71}, 071103 (2005).
\bibitem{CDFjpc} A. Abulencia {\it et al.} [CDF Collaboration], Phys. Rev. Lett. {\bf 98}, 132002 (2007).
\bibitem{rho} A. Abulencia {\it et al.} [CDF Collaboration], Phys. Rev. Lett. {\bf 96}, 102002 (2006).
\bibitem{Xomega} K. Abe {\it et al.} [Belle Collaboration], arXiv: hep-ex/ 0505037.
\bibitem{Belleddpi} G. Gokhroo {\it et al.} [Belle Collaboration], Phys. Rev. Lett. {\bf 97}, 162002 (2006).
\bibitem{Babarddpi} B. Aubert {\it et al.} [BABAR Collaboration], Phys. Rev. D {\bf 77}, 011102 (2008).
\bibitem{Belle3875v2} T. Aushev {\it et al.} [Belle Collaboration], Phys. Rev. D {\bf 81}, 031103 (2010).
\bibitem{recon} C. Hanhart, Yu. S. Kalashnikova, A. E. Kudryavtsev, and A. V. Nefediev, Phys. Rev. D {\bf 76}, 034007 (2007).
\bibitem{xnew} Yu. S. Kalashnikova and A. V. Nefediev, Phys. Rev. D {\bf 80}, 074004 (2009).
\bibitem{braaten} E. Braaten and J. Stapleton, Phys. Rev. D {\bf 81}, 014019 (2010).
\bibitem{babarjpc} P. del Amo Sanchez {\it et al.}, Phys. Rev. D {\bf 82}, 011101 (2010).
\bibitem{BG} T. Barnes and S. Godfrey, Phys. Rev. D {\bf 69}, 054008 (2004).
\bibitem{alla} A. M. Badalian, V. L. Morgunov, and B. L. G. Bakker, Phys. At. Nucl. {\bf 63}, 1635 (2000).
\bibitem{hch} T. Barnes, S. Godfrey, and E. S. Swanson, Phys. Rev. D {\bf 72}, 054026 (2005).
\bibitem{burns} T. J. Burns, F. Piccinini, A. D. Polosa, and C. Sabelli, Phys. Rev. D {\bf 82}, 074003 (2010).
\bibitem{eichten} E. J. Eichten, K. Lane, and C. Quigg, Phys. Rev. D {\bf 69}, 094019 (2004) 
\bibitem{yusk} Yu. S. Kalashnikova, Phys. Rev. D {\bf 72}, 034010 (2005).
\bibitem{chinese} Yu Jia, Wen-Long Sang, and Jia Xu, arXiv:1007.4541 [hep-ph].
\bibitem{babargamma} B. Aubert {\it et al} [BABAR Collaboration], Phys. Rev. Lett. {\bf 102}, 132001 (2009).
\bibitem{total} B. Aubert {\it et al} [BABAR Collaboration], Phys. Rev. Lett. {\bf 96}, 052002 (2006).
\bibitem{fazio} F. De Fazio, Phys. Rev. D {\bf 79}, 054015 (2009).
\bibitem{PDG} C. Amsler {\it et al.} (Particle Data Group), Phys. Lett. B {\bf 667}, 1 (2008).
\bibitem{bellegamma} V. Bhardwaj [Belle Collaboration], belle.kek.jp/talks/ QWQ2010/bhardwaj.pdf.
\bibitem{dunwoodie} W. Dunwoodie and V. Ziegler, Phys. Rev. Lett. {\bf 100}, 062006 (2008).
\bibitem{fw} M. Nauenberg and A. Pais, Phys. Rev. {\bf 126}, 360 (1962); 
E. Braaten and M. Lu, Phys. Rev. D {\bf 76}, 094028 (2007); 
C. Hanhart, Yu. S. Kalashnikova, A. V. Nefediev, Phys. Rev. D {\bf 81}, 031103 (2010).
\bibitem{loops} T. Barnes and E. S. Swanson, Phys. Rev. C {\bf 77}, 055206 (2008).
\end{thebibliography}
\end{document}